\documentclass[zpreprint,zbstpl]{./zeus_paper}
\usepackage[english]{babel}
\chardef\usc=95
\chardef\til=126
\catcode`\@=11 
\DeclareRobustCommand\xdotspace{\futurelet\@let@token\@xdotspace}
\def\@xdotspace{%
  \ifx\@let@token.\else
  \ifx\@let@token\bgroup.\else
  \ifx\@let@token\egroup.\else
  \ifx\@let@token\/.\else
  \ifx\@let@token\ .\else
  \ifx\@let@token~.\else
  \ifx\@let@token!.\else
  \ifx\@let@token,.\else
  \ifx\@let@token:.\else
  \ifx\@let@token;.\else
  \ifx\@let@token?.\else
  \ifx\@let@token/.\else
  \ifx\@let@token'.\else
  \ifx\@let@token).\else
  \ifx\@let@token-.\else
  \ifx\@let@token\@xobeysp.\else
  \ifx\@let@token\space.\else
  \ifx\@let@token\@sptoken.\else
   .\space
   \fi\fi\fi\fi\fi\fi\fi\fi\fi\fi\fi\fi\fi\fi\fi\fi\fi\fi}
\catcode`\@=12 

\newcommand{\stru}[2]{%
   \relax\ifmmode\hbox{\vrule height#1 depth#2 width0pt}%
   \else\vrule height#1 depth#2 width0pt\fi}

\newcommand{\Ronum}[1]{\uppercase\expandafter{\romannumeral#1}}
\newcommand{\ronum}[1]{\expandafter{\romannumeral#1}}
\DeclareRobustCommand{\LaTeXZ}{%
  \LaTeX\kern-.05em4\kern-.1em
  {\raisebox{-0.2ex}{$\scriptstyle\text{ZEUS}$}}\xspace}

\DeclareMathAlphabet{\mathbf}{OT1}{cmr}{bx}{sl}
\newcommand{\eVdist}{\kern-0.06667em}

\newcommand{\pb}{\,\text{pb}}

\newcommand{\slashfrac}[2]{%
  \raisebox{0.5ex}{\ensuremath #1}\kern-0.12em/\kern-0.08em
  \raisebox{-.8ex}{\ensuremath #2}}

\newcommand{\sqr}[3]{%
    {\vcenter{\hrule height.#3ex\hbox{\vrule width.#2ex height#1ex
     \kern#1ex\vrule width.#3ex}\hrule height.#2ex}}}

\catcode`\@=11 
\newcommand{\parenbar}{\mathpalette\p@renb@r}
\def\p@renb@r#1#2{\vbox{%
  \ifx#1\scriptscriptstyle \dimen@.7em\dimen@ii.2em\else
  \ifx#1\scriptstyle \dimen@.8em\dimen@ii.25em\else
  \dimen@1em\dimen@ii.4em\fi\fi \offinterlineskip
  \ialign{\hfill##\hfill\cr
    \vbox{\hrule width\dimen@ii}\cr
    \noalign{\vskip-.3ex}%
    \hbox to\dimen@{$\mathchar300\hfil\mathchar301$}\cr
    \noalign{\vskip-.3ex}%
    $#1#2$\cr}}}
\catcode`\@=12 

\newcommand{\IP}{{\rm I$\kern-0.01667em$P}\xspace}

\mathchardef\qsm=63
\mathchardef\pls=43
\mathchardef\mns=512
\mathchardef\plm=518
\mathchardef\eql=61
\mathchardef\smallleft=300
\mathchardef\smallright=301
\mathchardef\les=316
\mathchardef\gre=318
\mathchardef\leq=532
\mathchardef\grq=533

\catcode`\@=11 
\newcounter{pict@width}
\newcounter{pict@height}
\newlength{\pict@scale}
\newcommand{\psfigadd}[4]{%
\setcounter{pict@width}{1*\ratio{#2+\pict@scale/2}{\pict@scale}}
\setcounter{pict@height}{1*\ratio{#3+\pict@scale/2}{\pict@scale}}
\setlength{\unitlength}{\pict@scale}
\hbox to #2{\hspace{-\fill}\begin{picture}(\thepict@width,\thepict@height)
\put(0,0){\psfig{figure=#1,width=#2,height=#3,clip=}}
\SetScale{0.283466457}
\SetWidth{1.763889}
{#4}
\end{picture}}
}
\newcounter{pict@widthfst}
\newcounter{pict@widthscd}
\newcounter{pict@widthtot}
\newcommand{\psfigaddtwo}[7]{%
\setcounter{pict@widthfst}{1*\ratio{#2+\pict@scale/2}{\pict@scale}}
\setcounter{pict@widthscd}{1*\ratio{#2+#4+\pict@scale/2}{\pict@scale}}
\setcounter{pict@widthtot}{1*\ratio{#2+#4+#6+\pict@scale/2}{\pict@scale}}
\setcounter{pict@height}{1*\ratio{#3+\pict@scale/2}{\pict@scale}}
\setlength{\unitlength}{\pict@scale}
\hbox{\hspace{-\fill}\begin{picture}(\thepict@widthtot,\thepict@height)
\put(0,0){\psfig{figure=#1,width=#2,height=#3,clip=}}
\put(\thepict@widthscd,0){\psfig{figure=#5,width=#6,height=#3,clip=}}
\SetScale{0.283466457}
\SetWidth{1.763889}
{#7}
\end{picture}}
}
\newcommand{\psfigror}[4]{%
\setcounter{pict@width}{1*\ratio{#2+\pict@scale/2}{\pict@scale}}
\setcounter{pict@height}{1*\ratio{#3+\pict@scale/2}{\pict@scale}}
\setlength{\unitlength}{\pict@scale}
\hbox{\begin{picture}(\thepict@width,\thepict@height)
\put(0,\thepict@height){\psfig{figure=#1,width=#3,height=#2,clip=,angle=270}}
\SetScale{0.283466457}
\SetWidth{1.763889}
{#4}
\end{picture}}
}
\newcommand{\psfigrol}[4]{%
\setcounter{pict@width}{1*\ratio{#2+\pict@scale/2}{\pict@scale}}
\setcounter{pict@height}{1*\ratio{#3+\pict@scale/2}{\pict@scale}}
\setlength{\unitlength}{\pict@scale}
\hbox{\begin{picture}(\thepict@width,\thepict@height)
\put(0,0){\psfig{figure=#1,width=#3,height=#2,clip=,angle=90}}
\SetScale{0.283466457}
\SetWidth{1.763889}
{#4}
\end{picture}}
}
\catcode`\@=12 
\newlength\listtextwidth

\catcode`\@=11 
\newlength{\@tabfninsert}
\newlength{\@tabfnwidth}
\newcommand{\tabfootnote}[2]{%
  \setlength{\@tabfninsert}{0.8em}
  \setlength{\@tabfnwidth}{\textwidth}
  \addtolength{\@tabfnwidth}{-\@tabfninsert}
  \addtolength{\@tabfnwidth}{-0.4em}
  \noindent\makebox[\@tabfninsert][r]{\footnotesize$^{#1}$\hfil}\hfill%
  \parbox[t]{\@tabfnwidth}{\footnotesize #2\hfill}}
\catcode`\@=12 

\def\pb1{pb$^{-1}$}

\def\kt{k_T}

\def\etjetb{E^B_{T,\rm{jet}}}
\def\etajetb{\eta^B_{\rm{jet}}}
\def\phijetb{\phi^B_{\rm{jet}}}
\def\etjetl{E^L_{T,\rm{jet}}}
\def\etajetl{\eta^L_{\rm{jet}}}

\def\etabr{-2 < \etajetb < 1.8}

\def\setab{d\sigma/d\etajetb}

\def\setb{d\sigma/d\etjetb}
\def\sq2{d\sigma/dQ^2}

\def\ptmiss{p_T\hspace{-4.2mm}\slash\hspace{1.5mm}}
\def\etaphi{\eta-\phi}

\def\etib{E^B_{T,i}}
\def\etaib{\eta^B_i}
\def\phiib{\phi^B_i}
\def\etjb{E^B_{T,j}}
\def\etajb{\eta^B_j}
\def\phijb{\phi^B_j}

\def\cogam{\cos{\gamma}}
\def\cogamr{-0.7 < \cogam < 0.5}

\def\citeZEUS{{\cite{%
pl:b293:465,zeus:1993:bluebook%
}}\xspace}     
\def\citeCTD{{\cite{%
nim:a279:290,*npps:b32:181,*nim:a338:254%
}}\xspace}
\def\citeCAL{{\cite{%
nim:a309:77,*nim:a309:101,*nim:a321:356,*nim:a336:23%
}}\xspace}

\def\calscale{pl:b531:9,*epj:c23:615,*hepex0206036}

\def\citeLUMIMON{{\cite{%
desy-92-066,*zfp:c63:391,*actaphyspol:b32:2025%
}}\xspace}   

\def\disent{np:b485:291}
\def\disaster{graudenz:1997,*hepph9710244}
 
\def\citeKT{{\cite{%
np:b406:187%
}}\xspace}
 
\def\citeKTes{{\cite{%
pr:d48:3160%
}}\xspace}

\def\bwebber{jp:g19:1567}

\def\klasen{pl:b366:385,*np:b507:315,*cpc:133:105}

\def\papelshapes{epj:c8:367}
\def\papelhighqcuad{epj:c11:427}

\def\papelas{pl:b507:70}
\def\papeldijetdis{epj:c23:13}

\def\sinistra{nim:a365:508}
 
\def\dameth{proc:hera:1991:23}
 
\def\snow{proc:snowmass:1990:134}
 
\def\lepto{cpc:101:108}
 
\def\heracles{cpc:69:155,*spi:www:heracles}
 
\def\django{cpc:81:381,*spi:www:djangoh11}
 
\def\cdm{pl:b165:147,*pl:b175:453,*np:b306:746,*zfp:c43:625}
\def\ariadne{cpc:71:15,*zp:c65:285}
 
\def\lund{prep:97:31}
\def\jetset{cpc:39:347,*cpc:43:367}
 
\def\clustering{np:b238:492}
\def\herwig63{cpc:67:465,*jhep:0101:010,*hepph0107071}
 
\def\dglap{sovjnp:15:438,*sovjnp:20:94,*parisi:1976,*jetp:46:641,*np:b126:298}

\def\fixed{pl:b282:475,*pl:b223:485,*pl:b237:592,*np:b483:3,*np:b487:3,*pr:d54:3006,*prl:79:1213,*prl:80:3715}

\def\collider{*np:b470:3,*epj:c21:33,*zfp:c72:399,*prl:81:5754,epj:c21:443}

\def\pdg{epj:c15:1}

\def\cteqfour{pr:d55:1280}

\def\cteqsix{hepph0201195}

\def\grv{zfp:c67:433,*epj:c5:461}

\def\mrst2001{epj:c23:73}
\def\mrstnininine{epj:c4:463,*epj:c14:133}  

\def\botje{epj:c14:285}

\def\bethkeas{jp:g26:r27}

\begin{document}
\prepnum{{DESY--02--112}}      

\title{
Inclusive jet cross sections in the Breit\\
frame in neutral current deep inelastic\\
scattering at HERA and determination of $\alpha_s$   
}                                                       
                    
\author{ZEUS Collaboration}
\date{August, 2002}

\abstract{
     Inclusive jet differential cross sections 
     have
     been measured in neutral current deep inelastic $e^+p$ scattering for
     boson virtualities
     $Q^2>125$~GeV$^2$. The data were taken 
     using the ZEUS detector at HERA 
     and correspond to
     an integrated luminosity of $38.6$~\pb1. Jets were identified in
     the Breit frame using the longitudinally invariant $\kt$ cluster
     algorithm. Measurements of differential inclusive jet cross
     sections 
     are presented as functions of jet transverse energy  ($\etjetb$),
     jet pseudorapidity 
     and
     $Q^2$, for jets with $\etjetb>8$~GeV. Next-to-leading-order QCD calculations
     agree well with the measurements both at high $Q^2$ and high $\etjetb$.
    The value of
    $\alpha_s(M_Z)$, determined from an analysis of $d\sigma/dQ^2$ for
    $Q^2>500$~GeV$^2$, is
    $\alpha_s(M_Z) = 0.1212 \pm 0.0017 \; {\rm (stat.)}
                          ^{+ 0.0023}_{- 0.0031} \; {\rm (syst.)}
                          ^{+ 0.0028}_{-0.0027} \; {\rm (th.)}$.    
}

\makezeustitle

\def\3{\ss}                                                                                        
\pagenumbering{Roman}                                                                              
                                                   %
\begin{center}                                                                                     
{                      \Large  The ZEUS Collaboration              }                               
\end{center}                                                                                       
  S.~Chekanov,                                                                                     
  D.~Krakauer,                                                                                     
  S.~Magill,                                                                                       
  B.~Musgrave,                                                                                     
  J.~Repond,                                                                                       
  R.~Yoshida\\                                                                                     
 {\it Argonne National Laboratory, Argonne, Illinois 60439-4815}~$^{n}$                            
\par \filbreak                                                                                     
  M.C.K.~Mattingly \\                                                                              
 {\it Andrews University, Berrien Springs, Michigan 49104-0380}                                    
\par \filbreak                                                                                     
  P.~Antonioli,                                                                                    
  G.~Bari,                                                                                         
  M.~Basile,                                                                                       
  L.~Bellagamba,                                                                                   
  D.~Boscherini,                                                                                   
  A.~Bruni,                                                                                        
  G.~Bruni,                                                                                        
  G.~Cara~Romeo,                                                                                   
  L.~Cifarelli,                                                                                    
  F.~Cindolo,                                                                                      
  A.~Contin,                                                                                       
  M.~Corradi,                                                                                      
  S.~De~Pasquale,                                                                                  
  P.~Giusti,                                                                                       
  G.~Iacobucci,                                                                                    
  A.~Margotti,                                                                                     
  R.~Nania,                                                                                        
  F.~Palmonari,                                                                                    
  A.~Pesci,                                                                                        
  G.~Sartorelli,                                                                                   
  A.~Zichichi  \\                                                                                  
  {\it University and INFN Bologna, Bologna, Italy}~$^{e}$
% ...
\newpage                                                                                           
% ...
$^{\    2}$ on leave of absence at University of                                                   
Erlangen-N\"urnberg, Germany\\                                                                     
% ...
                                                           %
                                                           %
\newpage   
                                                           %
                                                           %
\begin{tabular}[h]{rp{14cm}}                                                                       
$^{a}$ &  supported by the Natural Sciences and Engineering Research                               
          Council of Canada (NSERC) \\                                                             
$^{b}$ &  supported by the German Federal Ministry for Education and                               
          Research (BMBF), under contract numbers HZ1GUA 2, HZ1GUB 0, HZ1PDA 5, HZ1VFA 5\\         
$^{c}$ &  supported by the MINERVA Gesellschaft f\"ur Forschung GmbH, the                          
          Israel Science Foundation, the U.S.-Israel Binational Science                            
          Foundation, the Israel Ministry of Science and the Benozyio Center                       
          for High Energy Physics\\                                                                
$^{d}$ &  supported by the German-Israeli Foundation, the Israel Science                           
          Foundation, and by the Israel Ministry of Science\\                                      
$^{e}$ &  supported by the Italian National Institute for Nuclear Physics (INFN) \\                
$^{f}$ &  supported by the Japanese Ministry of Education, Science and                             
          Culture (the Monbusho) and its grants for Scientific Research\\                          
$^{g}$ &  supported by the Korean Ministry of Education and Korea Science                          
          and Engineering Foundation\\                                                             
$^{h}$ &  supported by the Netherlands Foundation for Research on Matter (FOM)\\                   
$^{i}$ &  supported by the Polish State Committee for Scientific Research,                         
          grant no. 620/E-77/SPUB-M/DESY/P-03/DZ 247/2000-2002\\                                   
$^{j}$ &  partially supported by the German Federal Ministry for Education                         
          and Research (BMBF)\\                                                                    
$^{k}$ &  supported by the Fund for Fundamental Research of Russian Ministry                       
          for Science and Edu\-cation and by the German Federal Ministry for                       
          Education and Research (BMBF)\\                                                          
$^{l}$ &  supported by the Spanish Ministry of Education and Science                               
          through funds provided by CICYT\\                                                        
$^{m}$ &  supported by the Particle Physics and Astronomy Research Council, UK\\                   
$^{n}$ &  supported by the US Department of Energy\\                                               
$^{o}$ &  supported by the US National Science Foundation\\                                        
$^{p}$ &  supported by the Polish State Committee for Scientific Research,                         
          grant no. 112/E-356/SPUB-M/DESY/P-03/DZ 301/2000-2002, 2 P03B 13922\\                    
$^{q}$ &  supported by the Polish State Committee for Scientific Research,                         
          grant no. 115/E-343/SPUB-M/DESY/P-03/DZ 121/2001-2002, 2 P03B 07022\\                    
\end{tabular}                                                                                      
                                                           %
                                                           %

\pagenumbering{arabic} 
\pagestyle{plain}
%
\section{Introduction}
\label{sec-int}

  Jet production in neutral current deep inelastic $e^+ p$ scattering at high $Q^2$
(where $Q^2$ is the negative of the square of the virtuality of the exchanged
boson) provides a testing ground for the theory of the strong interaction
between quarks and gluons, namely quantum chromodynamics (QCD). In deep
inelastic scattering (DIS), the predictions of perturbative QCD (pQCD)
have the form of a convolution of matrix elements with parton
distribution functions (PDFs) of the target hadron. The matrix elements
describe the short-distance structure of the interaction and are calculable
in pQCD at each order, whilst the PDFs contain the description of the long-distance
structure of the target hadron.

 The evolution of the PDFs with the scale at which they are probed is predicted
in pQCD to follow a set of renormalisation group equations (DGLAP equations
\cite{\dglap}). However, an explicit determination of the PDFs requires
experimental input. A wealth of data from fixed-target \cite{\fixed} and
collider \cite{\collider} experiments has allowed an accurate determination
of the proton PDFs \cite{\cteqfour,\cteqsix,\grv,\mrstnininine,\mrst2001,\botje}. 
Good knowledge of PDFs makes measurements of jet production in DIS a sensitive
test of the pQCD predictions of the short-distance structure of the partonic interactions.
 
 The hadronic final state in neutral current DIS may consist of jets of high
transverse energy produced in the short-distance process as well as the remnant
(beam jet) of the incoming proton. A jet algorithm should distinguish as clearly
as possible between the beam jet and the hard jets. Working in the Breit
frame \cite{bookfeynam:1972,*zfp:c2:237} is preferred, since it provides a maximal
separation between the products of the beam fragmentation and the hard jets. In this frame,
the exchanged virtual boson ($V^{*}$, with $V^{*}=\gamma, Z$) is purely
space-like, with 3-momentum ${\bf q} = (0,0,-Q)$. In the Born process, the virtual
boson is absorbed by the struck quark, which is back-scattered with zero transverse
momentum with respect to the $V^{*}$ direction, whereas the beam jet follows the
direction of the initial struck quark. Thus, the contribution due to the current
jet in events from the Born process is suppressed by requiring the production of jets
with high transverse energy in this frame. Jet production in the Breit frame is,
therefore, directly sensitive to hard QCD processes, thus allowing direct tests of the
pQCD predictions. The use of the $\kt$ cluster algorithm~\citeKT to define jets in the
Breit frame facilitates the separation of the beam fragmentation and the hard process in
the calculations~\cite{\bwebber}.

At leading order (LO) in the strong coupling constant, $\alpha_s$, the boson-gluon-fusion
(BGF, $V^{*} g \rightarrow q\bar{q}$) and QCD-Compton (QCDC,
$V^{*} q \rightarrow qg$) processes give rise to two hard jets with opposite
transverse momenta. The calculation of dijet cross sections in pQCD at
fixed order in $\alpha_s$ is hampered by infrared-sensitive regions, so that additional
jet-selection criteria must be applied to make reliable predictions~\cite{\klasen}.
This complication is absent in the case of cross-section calculations for inclusive
jet production. 

 This paper presents measurements of several differential cross sections for the
inclusive production of jets with high transverse energy in the Breit frame. The
analysis is restricted to large values of $Q^2$, $Q^2 > 125$~GeV$^2$, and
the jets were selected according to their transverse energies and pseudorapidities in
the Breit frame; in the definition of the cross sections, no cut was applied to the
jets in the laboratory frame. The measurements are compared to next-to-leading-order
(NLO) QCD calculations \cite{\disent} using currently available parameterisations of
the proton PDFs. The jet selection used allows a reduction in the
theoretical uncertainty of the NLO QCD calculations with respect to those of dijet
production~\cite{\papeldijetdis,\papelas}. A QCD analysis of the inclusive jet cross
sections has been performed, which yields a more precise determination of $\alpha_s$
than was previously possible at
HERA~\cite{\papelas,pl:b363:201,mandyfit,pl:b346:415,*epj:c5:625,*epj:c6:575,epj:c19:289,*epj:c21:33}. 

\section{Experimental set-up}
\label{sec-exp}






The data sample used in this analysis was collected with the ZEUS detector at HERA and
corresponds to an integrated luminosity of $38.6 \pm 0.6$~\pb1. During 1996-1997, HERA
operated with protons of energy $E_p=820$~GeV and positrons of energy $E_e=27.5$~GeV.
The ZEUS detector is described in detail elsewhere \citeZEUS. The main components used
in the present analysis are the central tracking detector~\citeCTD, positioned in a
1.43~T solenoidal magnetic field, and the uranium-scintillator sampling calorimeter
(CAL)~\citeCAL. The tracking detector was used to establish an interaction vertex. The
CAL covers $99.7\%$ of the total solid angle. It is divided into three parts with a
corresponding division in the polar angle\footnote{The ZEUS coordinate system is a
right-handed Cartesian system, with the $Z$ axis pointing in the proton beam direction,
referred to as the ``forward direction'', and the $X$ axis pointing left towards the
centre of HERA. The coordinate origin is at the nominal interaction point. The
pseudorapidity is defined as $\eta=-\ln(\tan\frac{\theta}{2})$.}, $\theta$, as viewed
from the nominal interaction point: forward (FCAL, $2.6^{\circ}<\theta<36.7^{\circ}$),
barrel (BCAL, $36.7^{\circ}<\theta<129.1^{\circ}$), and rear (RCAL,
$129.1^{\circ}<\theta<176.2^{\circ}$). The smallest subdivision of the CAL is called a
cell. Under test-beam conditions, the CAL relative energy resolution is
$18\%/\sqrt{E(\text{GeV})}$ for electrons and $35\%/\sqrt{E(\text{GeV})}$ for hadrons.
Jet energies were corrected for the energy lost in inactive material, typically about
1~radiation length, in front of the CAL. The effects of uranium noise were minimised by
discarding cells in the inner (electromagnetic) or outer (hadronic) sections if they
had energy deposits of less than 60~MeV or 110~MeV, respectively. A three-level trigger
was used to select events online \cite{zeus:1993:bluebook}.        
 
The luminosity was measured using the Bethe-Heitler reaction 
$e^+p\rightarrow e^+\gamma p$~\citeLUMIMON. The resulting small-angle energetic photons
were measured by the luminosity monitor, a lead-scintillator calorimeter placed in the
HERA tunnel at $Z=-107$~m.
 
\section{Data selection and jet search}
\label{secsel}
 
Neutral current DIS events were selected offline using criteria similar to those
reported previously \cite{\papelshapes}. The main steps are briefly discussed below.
 
The scattered-positron candidate was identified from the pattern of energy deposits
in the CAL \cite{\sinistra}. The energy ($E_{e}^{\prime}$) and polar angle
($\theta_{e}$) of the positron candidate were determined from the CAL
measurements. The $Q^2$ variable was reconstructed from the double angle method
($Q^2_{DA}$) \cite{\dameth}, which uses $\theta_{e}$ and an angle $\gamma$ that
corresponds, in the quark-parton model, to the direction of the scattered quark. The
angle $\gamma$ was reconstructed from the CAL measurements of the hadronic final
state \cite{\dameth}. The following requirements were imposed on the data sample:
\begin{itemize}
\item a positron candidate of energy $E_{e}^{\prime}>10$~GeV. This cut ensured a high
     and well understood positron-finding efficiency and suppressed background from
     photoproduction events, in which the scattered positron escapes down the rear beampipe;
\item $y_e<0.95$, where $y_e=1-E_{e}^{\prime}(1-\cos{\theta_{e}})/(2 E_e)$.
      This condition removed events in which fake positron candidates were found
      in the FCAL;
\item the total energy not associated with the positron candidate within
      a cone of radius 0.7 units in the pseudorapidity-azimuth ($\etaphi$) plane around the
      positron direction should be less than $10\%$ of the positron energy.
      This condition removed photoproduction and DIS events in which part of a jet
      was falsely identified as the scattered positron;
\item for $30^{\circ}<\theta_{e}<140^{\circ}$, the fraction of the
      positron energy within a cone of radius 0.3 units in the $\etaphi$ plane
      around the positron direction should be larger than 0.9; for
      $\theta_{e}<30^{\circ}$, the cut was raised to 0.98. This
      condition removed events in which a jet was falsely identified as the
      scattered positron;
\item the vertex position along the beam axis should be in the range $-38<Z<32$~cm;
\item $38<(E-p_Z)<65$~GeV, where $E$ is the total energy as measured
      by the CAL, $E=\sum_iE_i$, and $p_Z$ is the $Z$-component of the
      vector ${\bf p}=\sum_i {E_i} \bf{r_i}$; in both cases the sum runs
      over all CAL cells, $E_i$ is the energy of the CAL cell $i$
      and ${\bf r_i}$ is a unit vector along the line joining the
      reconstructed vertex and the geometric centre of the cell $i$.
      This cut removed events with large initial-state radiation and further
      reduced the background from photoproduction;      
\item $\ptmiss/\sqrt{E_T}<2.5$~GeV$^{1/2}$, where $\ptmiss$ is the
      missing transverse momentum as measured with the CAL
      ($\ptmiss\equiv\sqrt{p_X^2+p_Y^2}$) and $E_T$ is the
      total transverse energy in the CAL. This cut removed
      cosmic rays and beam-related background;
\item no second positron candidate with energy above 10~GeV and energy in the CAL,
      after subtracting that of the two positron candidates, below 4~GeV. This
      requirement removed elastic Compton scattering events ($ep \rightarrow e\gamma p$);
\item $Q^2_{DA}>125$~GeV$^2$;
\item $-0.7 < \cos{\gamma} < 0.5$. The lower limit avoided a region with limited
      acceptance due to the requirement on the energy of the scattered positron,
      whilst the upper limit was chosen to ensure good reconstruction of the
      jets in the Breit frame.   
\end{itemize}
 
The longitudinally invariant $\kt$ cluster algorithm \citeKT was used in the inclusive
mode~\citeKTes to reconstruct jets in the hadronic final state both in data and in Monte Carlo (MC)
simulated events (see Section~\ref{secmc}). In data, the algorithm was applied to the
energy deposits measured in
the CAL cells after excluding those associated with the scattered-positron candidate.
The jet search was performed in the pseudorapidity ($\eta^B$)-azimuth ($\phi^B$) plane
of the Breit frame. In the following discussion, $\etib$ denotes the transverse energy, 
$\etaib$ the pseudorapidity and $\phiib$ the azimuthal angle of object $i$ in the Breit
frame. For each pair of objects (where the initial objects are the energy deposits in
the CAL cells), the quantity
\begin{equation}
 d_{ij} = [(\etaib - \etajb)^2 + (\phiib - \phijb)^2 ] \cdot {\rm min}(\etib,\etjb)^2
\end{equation}
was calculated. For each individual object, the quantity $d_i = (\etib)^2$
was also calculated. If, of all the values $\{d_{ij},d_i \}$, $d_{kl}$ was
the smallest, then objects $k$ and $l$ were combined into a single new
object. If, however, $d_k$ was the smallest, then object $k$ was considered
a jet and was removed from the sample. The procedure was repeated until all
objects were assigned to jets. The jet variables were defined according to the
Snowmass convention \cite{\snow}:
\begin{equation}
 \etjetb = \sum_i \etib \; , \;
 \etajetb = \frac{\sum_i \etib \etaib}{\etjetb} \; , \;
 \phijetb = \frac{\sum_i \etib \phiib}{\etjetb}.
\end{equation}
This prescription was also used to determine the variables of the
intermediate objects.

After reconstructing the jet variables in the Breit frame, the massless four-momenta
were boosted into the laboratory frame, where the transverse energy~($E^L_{T,{\rm{jet}}}$),
the pseudorapidity~($\eta^L_{\rm{jet}}$) and the azimuthal angle~($\phi^L_{\rm{jet}}$) of
each jet were calculated. Energy corrections were
then applied to the jets in the laboratory frame and propagated into the jet transverse
energies in the Breit frame. In addition, the jet variables in the laboratory
frame were used to apply additional cuts on the selected sample: 
\begin{itemize}
\item events were removed from the sample if the distance of any of the jets to the
      positron candidate in the $\eta-\phi$ plane of the laboratory frame,
      \begin{equation} 
       d = \sqrt{ (\eta^L_{\rm{jet}}- \eta_e)^2 + (\phi^L_{\rm{jet}} - \phi_e)^2},
      \end{equation}
      was smaller than 1~unit. This requirement removed some background from
      photoproduction and improved the purity of the sample;
\item events were removed from the sample if any of the jets was in the backward
      region of the detector ($\eta^L_{\rm{jet}} < -2$). This requirement removed events
      in which a radiated photon from the positron was misidentified as a hadronic jet
      in the Breit frame;
\item jets with low transverse energy in the laboratory frame ($E^L_{T,\rm{jet}} < 2.5$~GeV)
      were not included in the final sample; this cut removed a small number of
      jets for which the uncertainty on the energy correction was large. 
\end{itemize} 
It should be noted that these cuts were applied to improve the efficiency and purity of
the sample of jets and were not used to define the phase-space region of the cross-section
measurements. The simulated events were used to correct these effects on the cross sections.
In particular, the effects of the last two cuts were estimated to be smaller than
$3\%$. The final data sample contained 8523~events with at least one jet satisfying
$\etjetb >8$~GeV and $\etabr$. With the above criteria, 5073~one-jet, 3262~two-jet, 182~three-jet
and 6~four-jet events were found. Since the net transverse momentum of the hadronic final state
in the Breit frame is zero, an event with a single jet, according to a given selection criterion,
must contain at least one other jet balancing its transverse momentum; however,
this jet will not
necessarily satisfy the jet-selection criteria.

\section{Monte Carlo simulation}
\label{secmc}
 
Samples of events were generated to determine the response of the detector to jets of
hadrons and the correction factors necessary to obtain the hadron-level jet cross sections.
The generated events were passed through the GEANT~3.13-based~\cite{tech:cern-dd-ee-84-1} ZEUS detector- and
trigger-simulation programs \cite{zeus:1993:bluebook}. They were reconstructed and analysed
by the same program chain as the data.
 
Neutral current DIS events were generated using the LEPTO~6.5 program \cite{\lepto}
interfaced to HERACLES~4.5.2 \cite{\heracles} via DJANGO~6.2.4 \cite{\django}. The HERACLES
program includes photon and $Z$ exchanges and first-order electroweak radiative
corrections. The QCD cascade was modelled with the colour-dipole model \cite{\cdm} by
using the ARIADNE~4.08 program \cite{\ariadne} and including the BGF process.
The colour-dipole model treats gluons emitted from quark-antiquark (diquark) pairs as
radiation from a colour dipole between two partons. This results in partons that are not
ordered in their transverse momenta. The CTEQ4D \cite{\cteqfour} proton PDFs were used.
As an alternative, samples of events were generated using the model of LEPTO based on
first-order QCD matrix elements plus parton showers (MEPS). For the generation of the
samples with MEPS, the option for soft-colour interactions was switched
off~\cite{epj:c11:251}. In both cases, fragmentation into hadrons was performed using the
LUND \cite{\lund} string model as implemented in JETSET~7.4 \cite{\jetset}.
 
The jet search was performed on the MC events using the energy measured in the CAL cells in
the same way as for the data. Using the sample of events generated with either ARIADNE or
LEPTO-MEPS and after applying the same offline selection as for the data, a good
description of the measured distributions for the kinematic and jet variables was found.   
The same jet algorithm was also applied to the hadrons (partons) to obtain the predictions
at the hadron (parton) level. The MC programs were used to correct the measured cross
sections for QED radiative effects.

\section{NLO QCD calculations}
\label{comnlo}
 
 The measurements were compared with NLO QCD (${\cal O}(\alpha_s^2)$) calculations
obtained using the program DISENT \cite{\disent}. The calculations were performed      
in the $\overline{MS}$ renormalisation and factorisation schemes using a generalised
version \cite{\disent} of the subtraction method \cite{np:b178:421}.
The number of flavours was set to five and the renormalisation ($\mu_R$) and
factorisation ($\mu_F$) scales were chosen to be $\mu_R= \etjetb$ and $\mu_F=Q$,
respectively. The strong coupling constant, $\alpha_s$, was calculated at two loops
with $\Lambda^{(5)}_{\overline{MS}}=220$~MeV, corresponding to $\alpha_s(M_{Z})=0.1175$.
The calculations were performed using the MRST99~\cite{\mrstnininine} parameterisations of
the proton PDFs. The jet algorithm described in Section~\ref{secsel} was also applied
to the partons in the events generated by DISENT in order to compute the
jet cross-section predictions. The results obtained with DISENT were cross-checked by
using the program DISASTER++~\cite{\disaster}. The differences were always within $2\%$ and
typically smaller than $1\%$~\cite{thesisog}.
 
Since the measurements refer to jets of hadrons, whereas the NLO QCD calculations
refer to partons, the predictions were corrected to the hadron level using
the MC models. The multiplicative correction factor ($C_{\rm had}$) was defined as the ratio
of the cross section for jets of hadrons over that for jets of partons, estimated by using
the MC programs described in Section~\ref{secmc}. In order to estimate the
uncertainty in the simulation of the fragmentation process, events were also generated
using the HERWIG 6.3~\cite{\herwig63} program, where the hadronisation is simulated
by using a cluster model~\cite{\clustering}. The mean of the ratios obtained with ARIADNE,
LEPTO-MEPS and HERWIG was taken as the value of $C_{\rm had}$, since the three predictions were
in good agreement. The value of $C_{\rm had}$ differs from unity by less than $10\%$, except
in the backward region of the Breit frame where it differs by $20\%$.
 
The NLO QCD predictions were also corrected for the $Z$-exchange contribution
by using LEPTO. The multiplicative correction factor was defined as the ratio of the cross
section for jets of partons obtained with both photon and $Z$ exchange over that
obtained with photon exchange only. The correction is negligible for $Q^2 < 2000$~GeV$^2$
but reaches $17\%$ in the highest-$Q^2$ region.

Several sources of uncertainty in the theoretical predictions were considered:
\begin{itemize}
  \item the uncertainty on the NLO QCD calculations due to terms beyond NLO, estimated by
        varying $\mu_R$ between $\etjetb/2$ and $2\etjetb$, was $\sim \pm 5\%$;
  \item the uncertainty on the NLO QCD calculations due to that on $\alpha_s (M_Z)$ was
        estimated by repeating the calculations using two additional sets of proton PDFs,
        MRST99$\uparrow\uparrow$ and MRST99$\downarrow\downarrow$~\cite{\mrstnininine}, 
        determined assuming $\alpha_s (M_Z)=0.1225$ and $0.1125$, respectively.
        The difference between the calculations using these sets and MRST99 was
        scaled by a factor of $60\%$ to reflect the current uncertainty on the world average
        of $\alpha_s$~\cite{\bethkeas}. The resulting uncertainty in the cross sections was
        $\sim \pm 5\%$;
  \item the variance of the hadronisation corrections as predicted by ARIADNE, LEPTO-MEPS and
        HERWIG was taken as the uncertainty in this correction, which was typically less than
        $1\%$;
  \item the uncertainty on the NLO QCD calculations due to the statistical and correlated
        systematic experimental uncertainties of each data set used in the determination
        of the proton PDFs was calculated, making use of the results of an
        analysis~\cite{\botje} that provided the covariance matrix of the fitted PDF
        parameters and the derivatives as a function of Bjorken $x$ and $Q^2$.
        The resulting uncertainty in the cross sections was typically $3\%$, reaching $5\%$ in the
        high-$\etjetb$ region. To estimate the uncertainties on the 
        cross sections due to the theoretical uncertainties affecting the extraction of the
        proton PDFs, the calculation of all the differential cross sections was repeated
        using a number of different parameterisations obtained under different theoretical
        assumptions in the DGLAP fit~\cite{\botje}. This uncertainty in the cross sections was
        typically $3\%$. 
\end{itemize}
 
The total theoretical uncertainty was obtained by adding in quadrature the
individual uncertainties listed above.
 

\section{Systematic uncertainties}
 
The following sources of systematic uncertainty were considered for the measured jet 
cross sections~\cite{thesisog,thesishr}:
\begin{itemize}
  \item the uncertainty in the absolute energy scale of the jets was estimated to be $\pm 1\%$
        for $\etjetl > 10$~GeV \cite{\calscale} and $\pm 3\%$ for lower $\etjetl$ values. The
        resulting uncertainty was $\sim 5\%$;
  \item the uncertainty in the absolute energy scale of the positron candidate was estimated to
        be $\pm 1\%$~\cite{epj:c21:443}. The resulting uncertainty was less than $1\%$; 
  \item the differences in the results obtained by using either ARIADNE or LEPTO-MEPS to
        correct the data for detector and QED effects were taken to represent systematic
        uncertainties. The uncertainty was typically smaller than $3\%$;      
  \item the analysis was repeated using an alternative technique~\cite{\papelhighqcuad} to
        select the scattered-positron candidate. The uncertainty was less than $2\%$;   
  \item the $\etjetl$ cut was raised to 4~GeV. The uncertainty was smaller than $1\%$;
  \item the cut in $\etajetl$ used to suppress the contamination due to photons falsely
        identified as jets in the Breit frame was set to $-3$ and to $-1.5$. The uncertainty
        was typically $\sim 1\%$;
  \item the uncertainty in the cross sections due to that in the simulation of the trigger and
        in the cuts used to select the data was typically less than $3\%$.
\end{itemize}
 
In addition, there was an overall normalisation uncertainty of $1.6\%$ from the luminosity
determination, which was not considered in the cross-section calculation.
 
The systematic uncertainties not associated with the absolute energy scale of the jets were
added in quadrature to the statistical uncertainties and are shown on the figures as error bars.
The uncertainty due to the absolute energy scale of the jets is shown separately as a shaded
band in each figure, due to the large bin-to-bin correlation.  


\section{Inclusive jet differential cross sections}
\label{secres}
 
The differential inclusive jet cross sections were measured in the kinematic region
$Q^2 > 125$~GeV$^2$ and $\cogamr$. These cross sections include every jet of hadrons in
the event with $\etjetb > 8$~GeV and $\etabr$ and were corrected for detector and QED radiative
effects.
 
 The measurements of the differential inclusive jet cross sections as functions of $Q^2$,
$\etjetb$ and $\etajetb$ are presented in Figs.~\ref{fig2}$-$\ref{fig4}
and in Tables~\ref{tablaq2}$-$\ref{tablaeta}.
The data points are plotted at the weighted mean in each bin of the corresponding variable
as predicted by the NLO QCD calculation. The measured $\sq2$ ($\setb$) exhibits a steep
fall-off over five (three) orders of magnitude in the $Q^2$ ($\etjetb$) range considered. In
the low-$Q^2$ region ($125 < Q^2 < 250$~GeV$^2$), the selected data sample covers
$3\cdot 10^{-3} < x < 2 \cdot 10^{-2}$, whereas in the high-$Q^2$ region ($Q^2 > 5000$~GeV$^2$),
the range is $ 6\cdot 10^{-2} < x < 0.25$.
 
 The measurements of the differential cross-section $\setb$ in different regions of $Q^2$ are
presented in Fig.~\ref{fig6} 
and in Tables~\ref{tablaetq21} and~\ref{tablaetq22}. 
The $\etjetb$ dependence of the cross section
becomes less steep as $Q^2$ increases.  

\section{Comparison to NLO QCD calculations}
 
\begin{figure}[p]
\setlength{\unitlength}{1.0cm}
\begin{picture} (19.0,18.50)
\put (0.0,0.0){\epsfig{figure=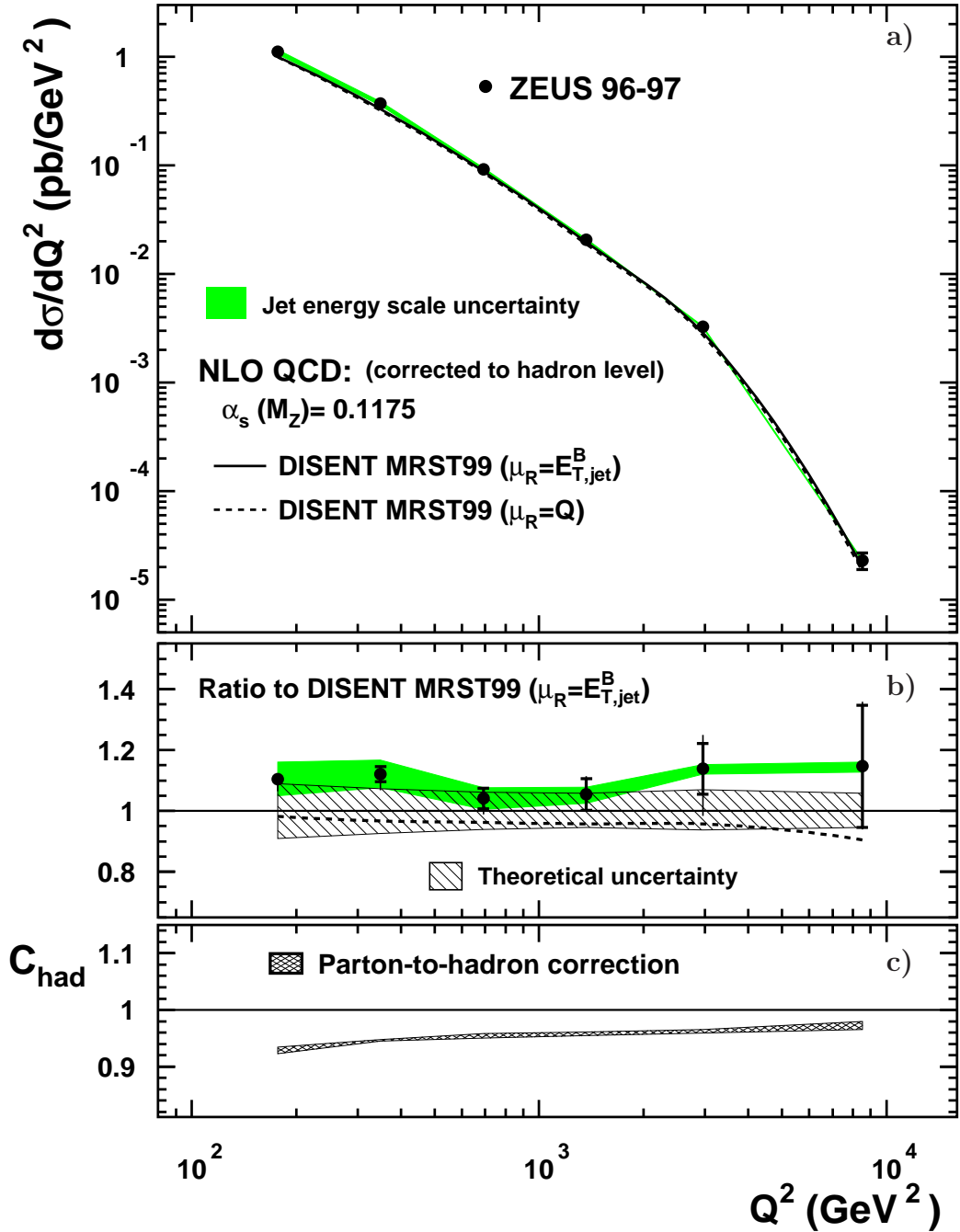,width=16.5cm}}
\put (13.0,18.4){{\bf a)}}
\put (13.0,9.0){{\bf b)}}
\put (13.0,5.0){{\bf c)}}
\end{picture}
\vspace{-2.0cm}
\caption{\label{fig2}
{a) The differential cross-section $\sq2$ for inclusive jet production
with $\etjetb > 8$~GeV and $\etabr$ (filled dots). The inner error bars represent
the statistical uncertainty. The outer error bars show the statistical and
systematic uncertainties, not associated with
the uncertainty in the absolute energy scale of the jets, added in
quadrature. The shaded band displays the uncertainty due to the absolute
energy scale of the jets. The NLO QCD calculations,
corrected for hadronisation effects and using the MRST99
parameterisations of the proton PDFs, are shown for two choices of the
renormalisation scale. b) The ratio between the measured
$\sq2$ and the NLO QCD calculation; the hatched band displays the
total theoretical uncertainty. The shaded band in c) shows the magnitude and
the uncertainty of the parton-to-hadron correction used to correct the NLO QCD
predictions.
}}          
\end{figure}

\begin{figure}[p]
\setlength{\unitlength}{1.0cm}
\begin{picture} (19.0,21.0)
\put (0.0,0.0){\epsfig{figure=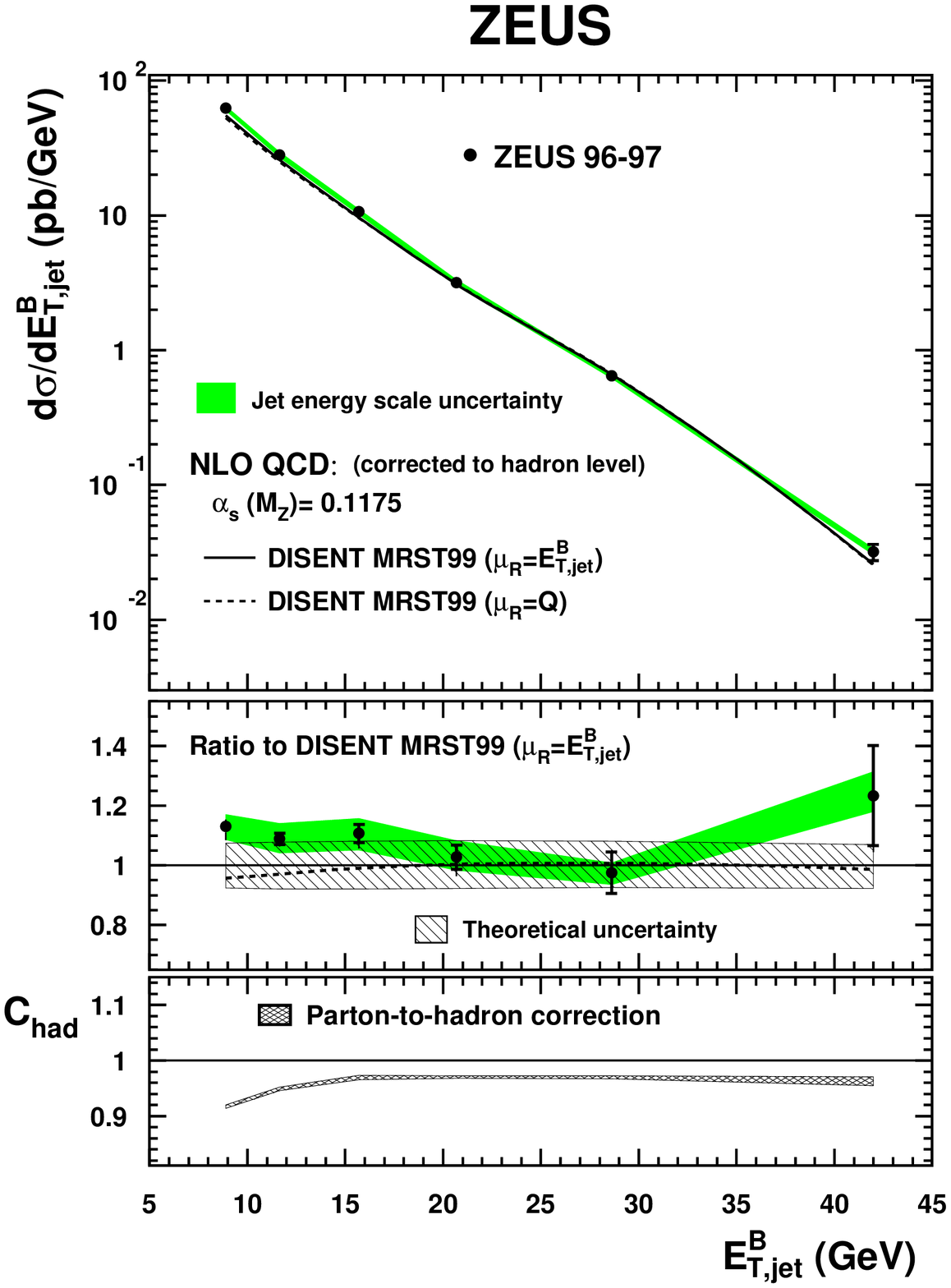,width=18.0cm}}
\put (14.20,20.0){{\bf a)}}
\put (14.20,10.){{\bf b)}}
\put (14.20,5.5){{\bf c)}}
\end{picture}
\vspace{-2.0cm}
\caption{\label{fig3}
{a) The differential cross-section $\setb$ for inclusive jet production
with $\etjetb > 8$~GeV and $\etabr$ (filled dots).
Other details are as described in the caption to Fig.~\ref{fig2}.}}
\end{figure}

\begin{figure}[p]
\setlength{\unitlength}{1.0cm}
\begin{picture} (19.0,21.0)
\put (0.0,0.0){\epsfig{figure=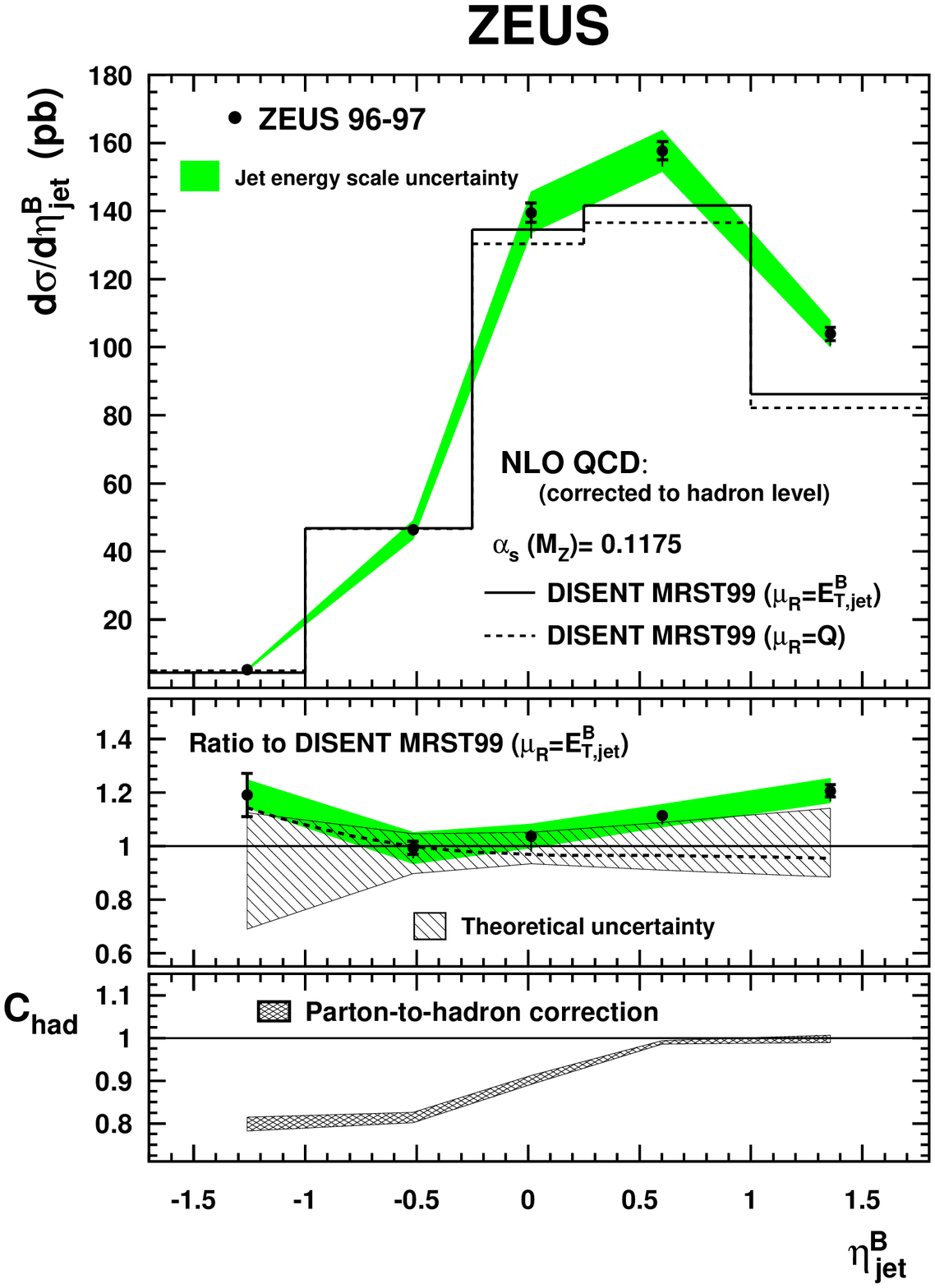,width=18cm}}
\put (14.30,20.0){{\bf a)}}
\put (14.30,10.){{\bf b)}}
\put (14.30,5.5){{\bf c)}}
\end{picture}
\vspace{-2.0cm}
\caption{\label{fig4}
{a) The differential cross-section $\setab$ for inclusive jet production
with $\etjetb > 8$~GeV and $\etabr$ (filled dots).
Other details are as described in the caption to Fig.~\ref{fig2}.}}
\end{figure}

\begin{figure}
\centerline{\hspace{2cm} \epsfig{figure=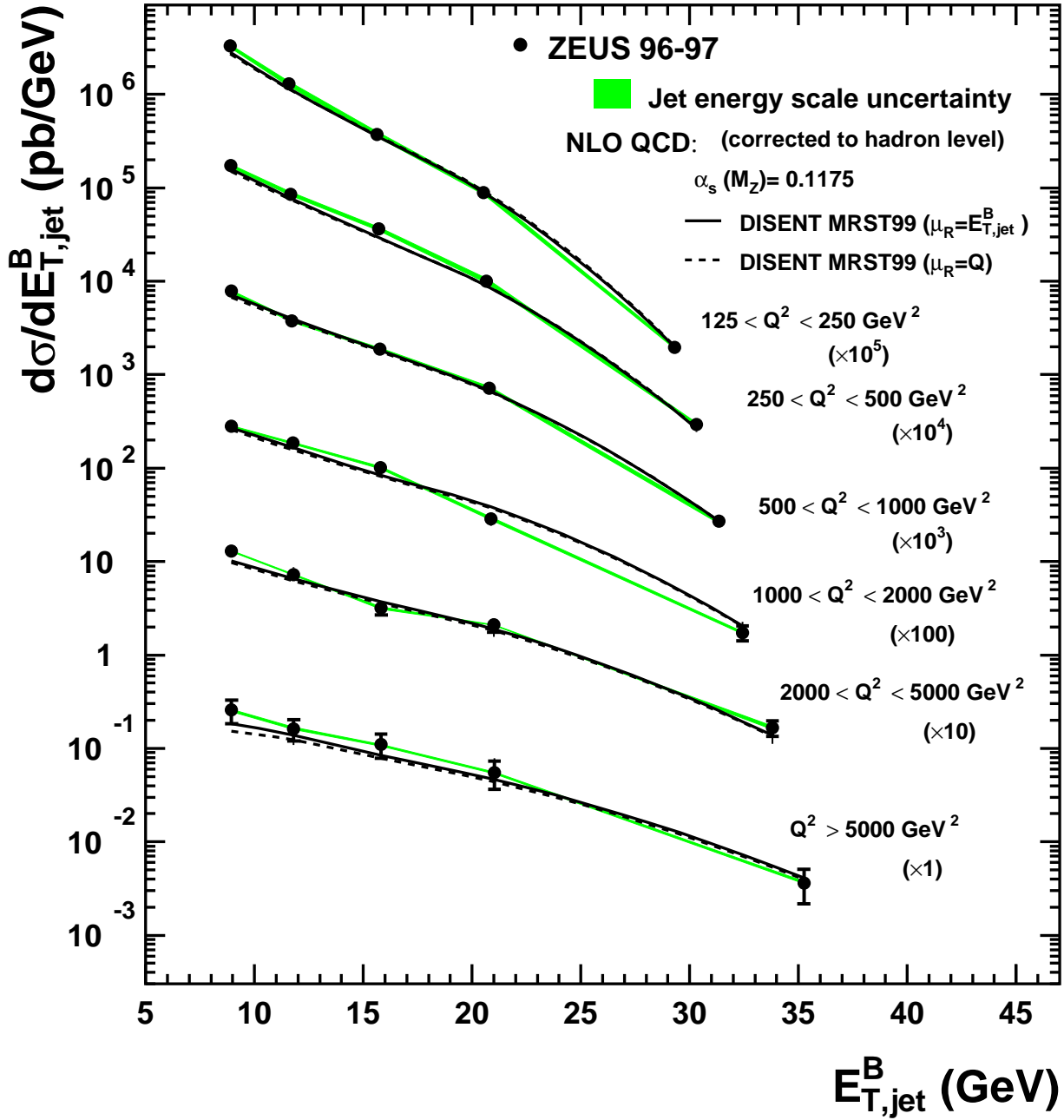,width=20cm}}
\vspace{-0.50cm}
\caption{\label{fig6}
{The differential cross-section $\setb$ for inclusive jet production
with $\etjetb > 8$~GeV and $\etabr$ in different regions 
of $Q^2$ (filled dots). 
Each cross section has been multiplied by the scale factor indicated
in brackets to aid visibility.
Other details are as described in the caption to Fig.~\ref{fig2}.}}
\end{figure}

 The NLO QCD predictions, corrected as described in Section~\ref{comnlo}, are displayed and
compared to the measurements in Figs.~\ref{fig2}-\ref{fig6}. It should be noted that the
hadronisation correction, shown in Figs.~\ref{fig2}c), \ref{fig3}c) and \ref{fig4}c), was obtained
with models (ARIADNE, LEPTO-MEPS and HERWIG) that implement higher-order contributions in
an approximate way and, thus, their predictions do not constitute genuine fixed-order NLO QCD
calculations. This procedure for applying hadronisation corrections to the NLO QCD calculations
was verified by checking that the shapes of the calculated differential cross sections
were well reproduced by the model predictions at the parton level.

\begin{figure}
\centerline{\hspace{2cm} \epsfig{figure=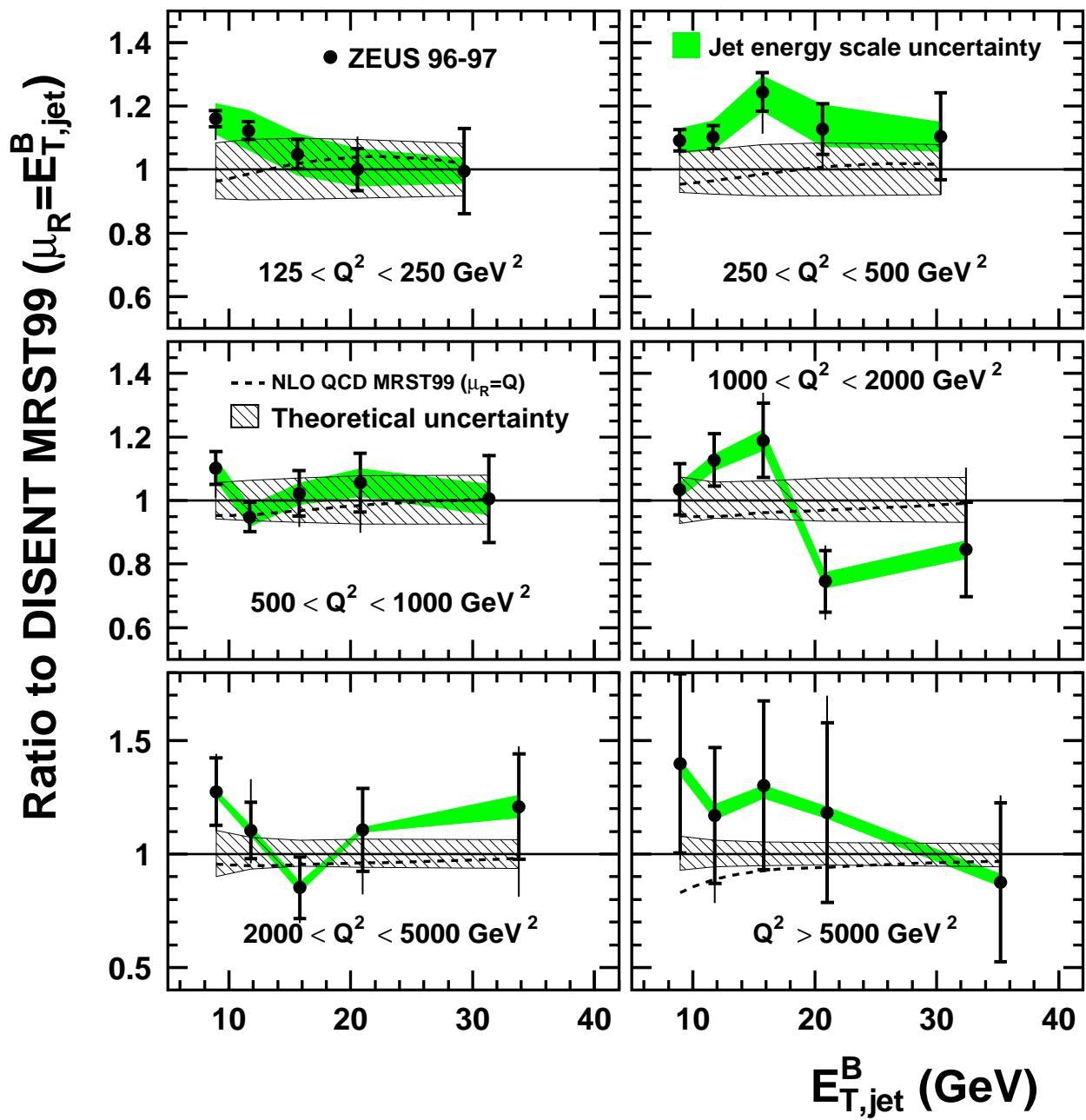,width=20cm}}
\vspace{-0.50cm}
\caption{\label{fig7}
{Ratios between the differential cross-sections $\setb$ presented in
Fig.~\ref{fig6} and NLO QCD calculations using the MRST99 parameterisations of
the proton PDFs and $\mu_R=\etjetb$ (filled dots).
Other details are as described in the caption to Fig.~\ref{fig2}.}}
\end{figure}

 The ratios of the measured differential cross sections over the NLO QCD calculations are
shown in Figs.~\ref{fig2}b), \ref{fig3}b), \ref{fig4}b) and \ref{fig7}. The calculations
reasonably reproduce the measured differential cross sections,
although they tend to be below the data. The agreement observed at high
$Q^2$ complements and extends an earlier comparison of the differential exclusive dijet
cross sections at $Q^2>470$~GeV$^2$~\cite{\papelas}. For that measurement of the exclusive
dijet cross sections, asymmetric cuts on the $\etjetb$ of the jets were applied \cite{\papelas} 
to avoid infrared-sensitive regions where NLO QCD programs are not reliable \cite{\klasen}.
This difficulty is not present in the calculations of inclusive jet cross sections and, as a
result, the theoretical uncertainties are smaller than in the dijet case. Thus, measurements
of inclusive jet cross sections allow more precise tests of the pQCD predictions than dijet
production.
 
 At low $Q^2$ and low $\etjetb$, the calculations fall below the data by $\sim 10\%$. The
differences between the measurements and calculations are of the same size as the
theoretical uncertainties. To study the scale dependence, NLO QCD calculations
using $\mu_R = \mu_F = Q$, shown as the dashed line, are also compared to the data
in Figs.~\ref{fig2}$-$\ref{fig7}; they provide a poorer description of the
data than those using $\mu_R=\etjetb$.
 
The overall description of the data by the NLO QCD calculations is sufficiently
good to make a precise determination of $\alpha_s$.
 
\section{Measurement of $\alpha_s$}
\label{secalphas}
 
The measured cross sections as a function of $Q^2$ and $\etjetb$ were used to
determine $\alpha_s(M_Z)$: 
\begin{itemize}
  \item NLO QCD calculations of $d\sigma/dA\;(A=Q^2,\etjetb)$ were performed for the
        three MRST99 sets, central, $\alpha_s\uparrow\uparrow$ and
        $\alpha_s\downarrow\downarrow$. The value of $\alpha_s(M_Z)$ used in each
        partonic cross-section calculation was that associated with the         
        corresponding set of PDFs;
  \item for each bin, $i$, in the variable $A$, the NLO QCD calculations,   
        corrected for hadronisation effects, were used to parameterise the
        $\alpha_s(M_Z)$ dependence of $d\sigma/dA$ according to
        \begin{equation}
        \label{e:fit_as}
 \left[ \frac{d\sigma}{dA} (\alpha_s(M_Z)) \right]_i = C_1^i\cdot \alpha_s(M_Z)+
        C_2^i\cdot {\alpha_s}^2(M_Z) \; ;
        \end{equation}
  \item the value of $\alpha_s(M_Z)$ was then determined by a $\chi^2$ fit
        of Eq.~(\ref{e:fit_as}) to the measured $d\sigma/dA$ values for
        several regions of the variable $A$.
\end{itemize}
 
This procedure correctly handles the complete $\alpha_s$ dependence of the NLO
differential cross sections (the explicit dependence coming from the partonic
cross sections and the implicit dependence coming from the PDFs) in the fit, while
preserving the correlation between $\alpha_s$ and the PDFs.
 
The uncertainty on the extracted values of $\alpha_s (M_Z)$ due to the
experimental systematic uncertainties was evaluated by repeating the analysis
above for each systematic check~\cite{thesisog}. The overall normalisation uncertainty
from the luminosity determination was also considered. The largest contribution
to the experimental uncertainty comes from the jet energy scale.  
 
The theoretical uncertainties, evaluated as described in Section~\ref{comnlo},
arising from terms beyond NLO, uncertainties in the proton PDFs and uncertainties in
the hadronisation correction were considered. These resulted in 
uncertainties in
$\alpha_s (M_Z)$ of $3\%$, $1\%$ and $0.2\%$, respectively. The total 
theoretical
uncertainty was obtained by adding these uncertainties in quadrature. 
The results are presented in Tables~\ref{tablaasq2} and~\ref{tablaaset}.

The best determination of $\alpha_s(M_Z)$ was obtained by using the measured $d\sigma/dQ^2$
for $Q^2>500$~GeV$^2$, for which both the theoretical and total uncertainties in $\alpha_s(M_Z)$
are minimised. A good fit was obtained with $\chi^2=2.1$ for 4~data points. The fitted value is
\begin{equation}                     
 \alpha_s(M_Z) = 0.1212 \pm 0.0017 \; {\rm (stat.)}
              ^{+ 0.0023}_{- 0.0031} \; {\rm (syst.)}
              ^{+ 0.0028}_{-0.0027} \; {\rm (th.)}\; . \nonumber  
\end{equation}
As a cross check, the measurement was repeated using the five sets of proton PDFs of the CTEQ4 
A-series~\cite{\cteqfour}; the result is in good agreement with the above value.
Two other determinations of $\alpha_s(M_Z)$ were performed. The first made
use of the measured $d\sigma/dQ^2$ for the entire $Q^2$ range studied, $Q^2>125$~GeV$^2$,
resulting in $\alpha_s(M_Z) = 0.1244 \pm 0.0009 \; {\rm (stat.)}
^{+ 0.0034}_{- 0.0041} \; {\rm(syst.)} ^{+ 0.0057}_{-0.0040} \; {\rm(th.)}$.
The second used the measured $d\sigma/d\etjetb$ in the region where the hadronisation corrections
are small, $\etjetb>14$~GeV, resulting in
$\alpha_s(M_Z) = 0.1212 \pm 0.0013 \; {\rm (stat.)}
 ^{+ 0.0030}_{- 0.0036} \; {\rm (syst.)} ^{+ 0.0041}_{-0.0030} \; {\rm (th.)}$.
These results are consistent with the central value quoted above.

The value of $\alpha_s(M_Z)$ obtained is consistent with the current PDG value,
$\alpha_s(M_Z)=0.1181 \pm 0.0020$~\cite{\pdg} and recent determinations by the
H1~\cite{epj:c19:289,*epj:c21:33} and ZEUS~\cite{\papelas,mandyfit} Collaborations. It is
compatible with a recent determination from the measurement of the inclusive jet cross
section in $p\bar{p}$ collisions at $\sqrt{s}=1800$~GeV,
$\alpha_s(M_Z)=0.1178 \pm 0.0001 {\rm (stat.)} ^{+0.0081} _{-0.0095} {\rm (syst.)}
^{+0.0092} _{-0.0075} {\rm (th.)}$ \cite{prl:88:042001}. It is in agreement 
with, and has a precision comparable to, the most accurate value 
obtained in $e^+e^-$ interactions~\cite{\bethkeas}.

\begin{figure}
\centerline{\hspace{2cm} \epsfig{figure=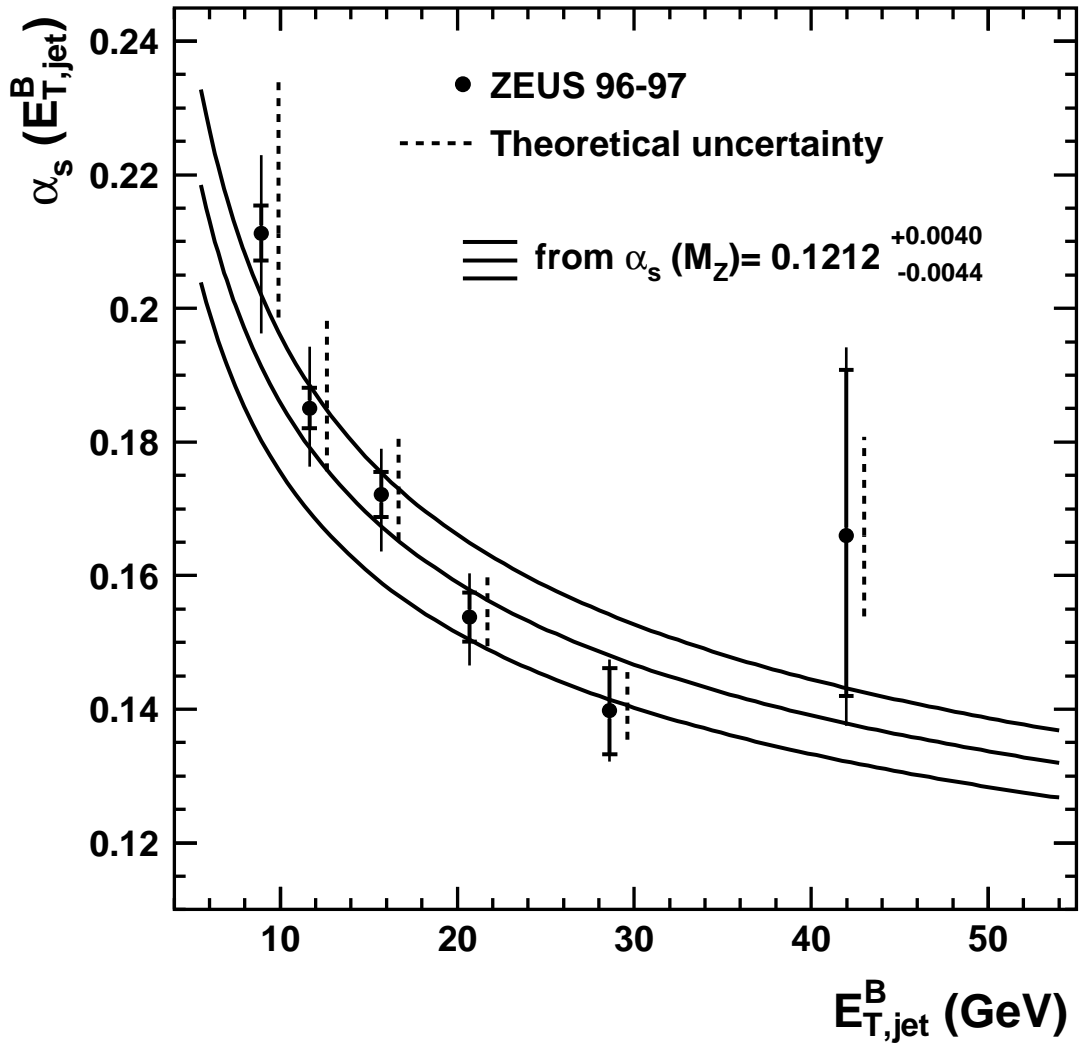,width=20cm}}
\vspace{-2.0cm}
\caption{\label{fig8}
{The $\alpha_s(\etjetb)$ values determined from the QCD fit of the 
measured $d\sigma/d\etjetb$ as a function of $\etjetb$. The 
inner error bars
represent the statistical uncertainty of the data. The outer error bars
show the statistical and systematic uncertainties added in quadrature.
The dashed error bars display the theoretical uncertainties. The three
curves indicate the renormalisation group predictions obtained from the
$\alpha_s(M_Z)$ central value determined in this analysis and its
associated uncertainty.}}
\end{figure}

The QCD prediction for the energy-scale dependence of the strong coupling constant has
been tested by determining $\alpha_s$ from the measured differential cross sections at
different scales. Since the NLO QCD calculations with $\mu_R=\etjetb$ provide a better
description of the data than those using $\mu_R=Q$, a QCD fit to the measured
$d\sigma/d\etjetb$ was performed in each bin of $\etjetb$. The principle of the fit is
the same as outlined above, with the only difference being
that the $\alpha_s$ dependence of $d\sigma/d\etjetb$ in Eq.~(\ref{e:fit_as}) was
parameterised in terms of $\alpha_s(\langle \etjetb \rangle)$ rather than $\alpha_s(M_Z)$,
where $\langle \etjetb \rangle$ is the
mean value of $\etjetb$ in each bin. The measured $\alpha_s(\langle \etjetb \rangle)$
values, with their experimental and theoretical systematic uncertainties estimated as for
$\alpha_s(M_{Z})$, are shown in Fig.~\ref{fig8}
and in Table~\ref{tablaaset2}. 
The measurements are compared with the
renormalisation group predictions obtained from the $\alpha_s(M_Z)$ central value determined
above and its associated uncertainty. The results are in good agreement with the
predicted running of the strong coupling constant over a large range in $\etjetb$.

\section{Summary}
\label{secsumm}
 
Measurements of the differential cross sections for inclusive jet production in neutral
current deep inelastic $e^+p$ scattering at a centre-of-mass energy of 300~GeV have
been presented. The cross sections refer to jets of hadrons identified with the
longitudinally invariant $\kt$ cluster algorithm in the Breit frame. The cross sections
are given in the kinematic region $Q^2> 125$~GeV$^2$ and $\cogamr$.
 
NLO QCD calculations provide a good description of the measured differential
cross sections for inclusive jet production at high $Q^2$, $Q^2 > 500$~GeV$^2$, or
high jet transverse energies, $\etjetb > 14$~GeV. This observation complements   
and extends that of the exclusive dijet cross section to lower $Q^2$. At low
$Q^2$ and low jet transverse energies, differences of $\sim 10\%$ between data and
calculations are observed, which are of the same size as the theoretical uncertainties.
 
A QCD fit of the measured cross section as a function of $Q^2$ for $Q^2 > 500$~GeV$^2$
yields
\begin{displaymath}
       \alpha_s(M_Z) = 0.1212 \pm 0.0017 \; {\rm (stat.)}
                          ^{+ 0.0023}_{- 0.0031} \; {\rm (syst.)}
                          ^{+ 0.0028}_{-0.0027} \; {\rm (th.)}.
\end{displaymath}
This value is in good agreement with the world average and is at least as precise as any other
individual measurement.
 
\newpage
\noindent {\Large\bf Acknowledgments}
\vspace{0.3cm}
 
We thank the DESY Directorate for their strong support and encouragement.
The remarkable achievements of the HERA machine group were essential for
the successful completion of this work and are greatly appreciated. We
are grateful for the support of the DESY computing and network services.
The design, construction and installation of the ZEUS detector have been
made possible owing to the ingenuity and effort of many people from DESY
and home institutes who are not listed as authors.

\vfill\eject

\raggedright
\providecommand{\etal}{et al.\xspace}
\providecommand{\coll}{Collaboration}
\catcode`\@=11
\def\@bibitem#1{%
\ifmc@bstsupport
  \mc@iftail{#1}%
    {;\newline\ignorespaces}%
    {\ifmc@first\else.\fi\orig@bibitem{#1}}
  \mc@firstfalse
\else
  \mc@iftail{#1}%
    {\ignorespaces}%
    {\orig@bibitem{#1}}%
\fi}%
\catcode`\@=12

\vfill\eject

\begin{table}[p]
\begin{center}
\mbox{
\hspace{-1.8cm}
\begin{tabular}{|c|cccc||c|c|}
\hline
\raisebox{0.25cm}[1.cm]{\parbox{2cm}{\centerline{$Q^2$ bin} \centerline{(GeV$^2$)}}}
          & \raisebox{0.25cm}[1.cm]{\parbox{2cm}{\centerline{$d\sigma/dQ^2$} \centerline{(pb/GeV$^2$)}}} 
                           & $\Delta_{stat}$ & 
                             $\Delta_{syst}$ & 
                             $\Delta_{\text{jet}-ES}$ & 
           \raisebox{0.2cm}[0.8cm]{\parbox{2.cm}{\centerline{QED} \vspace{-.2cm} \centerline{correction}}} &  
           \raisebox{0.2cm}[0.8cm]{\parbox{2.8cm}{\centerline{PAR to HAD} \vspace{-0.2cm} \centerline{correction}}}\\[.05cm]
\hline
   $125\;-\;250$ &
              $1.107$  & $\pm 0.018$ &  $^{+0.010}_{-0.035}$ &  $^{+0.056}_{-0.055}$ &
                           $0.950$ &  $0.9283 \pm 0.0058$
   \\
   $250\;-\;500$ & 
              $0.3714$  &  $\pm 0.0080$ & $^{+0.0038}_{-0.0153}$ & $^{+0.0156}_{-0.0148}$ &
                           $0.947$ & $0.9463 \pm 0.0014$
   \\
   $500\;-\;1000$ &
              $0.0919$ &  $\pm 0.0029$ & $^{+0.0008}_{-0.0035}$ & $^{+0.0031}_{-0.0032}$ &
                           $0.959$ & $0.9542 \pm 0.0038$
   \\
   $1000\;-\;2000$ &
              $0.02068$ &  $\pm 0.00103$ & $^{+0.00055}_{-0.00018}$ & $^{+0.00047}_{-0.00057}$ &
                           $0.955$ & $0.9579 \pm 0.0035$
   \\
   $2000\;-\;5000$ &
              $0.00325$ &  $\pm 0.00024$ & $^{+0.00021}_{-0.00037}$ & $^{+0.00004}_{-0.00005}$ &
                           $0.963$ & $0.9623 \pm 0.0028$
   \\
   $5000\;-\;10^{5}$ &
              \small $2.29\cdot 10^{-5}$ & \small $\pm 0.40\cdot 10^{-5}$ & $^{+0.14}_{-0.11} \cdot 10^{-5}$ & $^{+0.03}_{-0.04} \cdot 10^{-5}$ &
                           $0.918$ & $0.9727 \pm 0.0069$
   \\
\hline
\end{tabular}}
\caption{
Inclusive jet cross-section $d\sigma/dQ^2$ for jets of hadrons in the Breit frame,
selected with the longitudinally invariant $k_T$ cluster algorithm. The statistical, systematic and 
jet-energy-scale uncertainties are shown separately. The multiplicative correction
applied to correct for QED radiative effects and for hadronisation effects are shown
in the last two columns.
}
  \label{tablaq2}
\end{center}
\end{table}
\begin{table}[p]
\begin{center}
\begin{tabular}{|c|cccc||c|c|}
\hline
\raisebox{0.25cm}[1.cm]{\parbox{2cm}{\centerline{$\etjetb$ bin} \centerline{(GeV)}}}
                    & \raisebox{0.25cm}[1.cm]{\parbox{2cm}{\centerline{$d\sigma/d\etjetb$}  \centerline{(pb/GeV)}}}
                                   & $\Delta_{stat}$ & 
                                     $\Delta_{syst}$ & 
                                     $\Delta_{\text{jet}-ES}$ & 
            \raisebox{0.2cm}[0.8cm]{\parbox{2.cm}{\centerline{QED} \vspace{-.2cm} \centerline{correction}}} &  
           \raisebox{0.2cm}[0.8cm]{\parbox{2.8cm}{\centerline{PAR to HAD} \vspace{-0.2cm} \centerline{correction}}}\\[.05cm]
\hline
   $8\;-\;10$ &
              $62.42$  & $\pm 0.99$ & $^{+0.93}_{-2.35}$ & $^{+2.19}_{-2.39}$ &
                           $0.955$ & $0.9170 \pm 0.0030$
   \\
   $10\;-\;14$ & 
             $28.09$  & $\pm 0.49$ & $^{+0.23}_{-0.44}$ & $^{+1.33}_{-1.21}$ &
                           $0.951$ & $0.9488 \pm 0.0033$
   \\
   $14\;-\;18$ &
             $10.66$ & $\pm 0.29$ & $^{+0.05}_{-0.39}$ & $^{+0.49}_{-0.53}$ &
                           $0.955$ & $0.9697 \pm 0.0039$
   \\
   $18\;-\;25$ &
             $3.16$ & $\pm 0.12$ & $^{+0.04}_{-0.15}$ & $^{+0.17}_{-0.14}$ &
                           $0.954$ & $0.9703 \pm 0.0022$
   \\
   $25\;-\;35$ &
             $0.646$ & $\pm 0.046$ & $^{+0.020}_{-0.002}$ & $^{+0.022}_{-0.026}$ &
                           $0.944$ & $0.9698 \pm 0.0026$
   \\
   $35\;-\;100$ &
             $0.0318$ & $\pm 0.0043$ & $^{+0.0010}_{-0.0023}$ & $^{+0.0021}_{-0.0014}$ &
                           $0.954$ & $0.9627 \pm 0.0082$
   \\
\hline
\end{tabular}
\caption{
Inclusive jet cross-section $d\sigma/d\etjetb$ for jets of hadrons in the Breit frame,
selected with the longitudinally invariant $k_T$ cluster algorithm. For details, see the caption 
of Table~\ref{tablaq2}.
}
  \label{tablaet}
\end{center}
\end{table}
\begin{table}[p]
\begin{center}
\begin{tabular}{|c|cccc||c|c|}
\hline
\raisebox{0.15cm}[1.cm]{\parbox{2cm}{\centerline{$\etajetb$ bin}}} & 
\raisebox{0.25cm}[1.cm]{\parbox{2cm}{\centerline{$d\sigma/d\etajetb$} \centerline{(pb)}}} 
                                     & $\Delta_{stat}$ & 
                                       $\Delta_{syst}$ & 
                                       $\Delta_{\text{jet}-ES}$ & 
           \raisebox{0.2cm}[0.8cm]{\parbox{2.cm}{\centerline{QED} \vspace{-.2cm} \centerline{correction}}} &  
           \raisebox{0.2cm}[0.8cm]{\parbox{2.8cm}{\centerline{PAR to HAD} \vspace{-0.2cm} \centerline{correction}}}\\[.05cm]
\hline
   $-2\;-\; -1$ &
              $5.27$  & $\pm 0.36$ & $^{+0.09}_{-0.16}$ & $^{+0.25}_{-0.21}$ &
                           $0.942$ & $0.798 \pm 0.016$
   \\
   $-1\;-\; -0.25$ & 
             $46.5$  & $\pm 1.2$ & $^{+0.9}_{-1.4}$ & $^{+2.7}_{-2.8}$ &
                           $0.947$ & $0.813 \pm 0.012$
   \\
   $-0.25\;-\; 0.25$ &
             $139.5$ & $\pm 2.8$ & $^{+1.4}_{-7.1}$ & $^{+6.2}_{-6.0}$ &
                           $0.953$ & $0.901 \pm 0.010$
   \\
   $0.25 \;-\; 1$ &
             $157.7$ & $\pm 2.7$ & $^{+1.0}_{-3.9}$ & $^{+6.1}_{-6.1}$ &
                           $0.963$ & $0.9900 \pm 0.0040$
   \\
   $1 \;-\; 1.8$ &
             $103.9$ & $\pm 2.0$ & $^{+0.7}_{-2.8}$ & $^{+4.0}_{-3.8}$ &
                           $0.957$ & $0.9982 \pm 0.0088$
   \\
\hline
\end{tabular}
\caption{
Inclusive jet cross-section $d\sigma/d\etajetb$ for jets of hadrons in the Breit frame,
selected with the longitudinally invariant $k_T$ cluster algorithm. For details, see the caption 
of Table~\ref{tablaq2}.
}
  \label{tablaeta}
\end{center}
\end{table}
\begin{table}[p]
\begin{center}
\begin{tabular}{|c|cccc||c|c|}
\hline
\raisebox{0.25cm}[1.cm]{\parbox{2cm}{\centerline{$\etjetb$ bin} {\centerline{(GeV)}}}} & 
\raisebox{0.25cm}[1.cm]{\parbox{2cm}{\centerline{$d\sigma/d\etjetb$} \centerline{(pb/GeV)}}} & $\Delta_{stat}$ & 
                                     $\Delta_{syst}$ & 
                                     $\Delta_{\text{jet}-ES}$ & 
            \raisebox{0.2cm}[0.8cm]{\parbox{2.cm}{\centerline{QED} \vspace{-.2cm} \centerline{correction}}} &  
           \raisebox{0.2cm}[0.8cm]{\parbox{2.8cm}{\centerline{PAR to HAD} \vspace{-0.2cm} \centerline{correction}}}\\[.05cm]
\hline
\hline
\multicolumn{7}{|c|}{\raisebox{0.05cm}[0.5cm]{$\mathbf{125 < Q^2 < 250 \text{ \bf GeV}^2}$}} \\
\hline
   $8\;-\;10$ &
              $32.90$  & $\pm 0.71$ & $^{+0.48}_{-1.73}$ & $^{+1.35}_{-1.38}$ &
                           $0.965$ & $0.9137 \pm 0.0029$
   \\
   $10\;-\;14$ & 
             $13.02$  & $\pm 0.32$ & $^{+0.11}_{-0.37}$ & $^{+0.73}_{-0.69}$ &
                           $0.963$ & $0.9380 \pm 0.0075$
   \\
   $14\;-\;18$ &
             $3.75$ & $\pm 0.16$ & $^{+0.03}_{-0.06}$ & $^{+0.23}_{-0.24}$ &
                           $0.964$ & $0.9496 \pm 0.0069$
   \\
   $18\;-\;25$ &
             $0.895$ & $\pm 0.059$ & $^{+0.071}_{-0.015}$ & $^{+0.059}_{-0.047}$ &
                           $0.963$ & $0.9394 \pm 0.0041$
   \\
   $25\;-\;100$ &
             $0.0197$ & $\pm 0.0027$ & $^{+0.0002}_{-0.0000}$ & $^{+0.0008}_{-0.0008}$ &
                           $0.956$ & $0.9162 \pm 0.0067$
   \\
\hline
\hline
\multicolumn{7}{|c|}{\raisebox{0.05cm}[0.5cm]{$\mathbf{250 < Q^2 < 500 \text{ \bf GeV}^2}$}} \\
\hline
   $8\;-\;10$ &
              $17.33$  & $\pm 0.52$ & $^{+0.27}_{-0.54}$ & $^{+0.59}_{-0.65}$ &
                           $0.949$ & $0.9205 \pm 0.0084$
   \\
   $10\;-\;14$ & 
             $8.57$  & $\pm 0.28$ & $^{+0.08}_{-0.25}$ & $^{+0.39}_{-0.31}$ &
                           $0.942$ & $0.9573 \pm 0.0043$
   \\
   $14\;-\;18$ &
             $3.64$ & $\pm 0.18$ & $^{+0.04}_{-0.34}$ & $^{+0.15}_{-0.18}$ &
                           $0.953$ & $0.9748 \pm 0.0064$
   \\
   $18\;-\;25$ &
             $1.007$ & $\pm 0.072$ & $^{+0.009}_{-0.092}$ & $^{+0.068}_{-0.048}$ &
                           $0.952$ & $0.9685 \pm 0.0042$
   \\
   $25\;-\;100$ &
             $0.0294$ & $\pm 0.0036$ & $^{+0.0002}_{-0.0032}$ & $^{+0.0012}_{-0.0013}$ &
                           $0.937$ & $0.9539 \pm 0.0022$
   \\
\hline
\hline
\multicolumn{7}{|c|}{\raisebox{0.05cm}[0.5cm]{$\mathbf{500 < Q^2 < 1000 \text{ \bf GeV}^2}$}} \\
\hline
   $8\;-\;10$ &
              $7.83$  & $\pm 0.36$ & $^{+0.14}_{-0.55}$ & $^{+0.21}_{-0.26}$ &
                           $0.938$ & $0.9205 \pm 0.0090$
   \\
   $10\;-\;14$ & 
             $3.77$  & $\pm 0.18$ & $^{+0.10}_{-0.05}$ & $^{+0.14}_{-0.12}$ &
                           $0.941$ & $0.9579 \pm 0.0049$
   \\
   $14\;-\;18$ &
             $1.87$ & $\pm 0.13$ & $^{+0.02}_{-0.14}$ & $^{+0.06}_{-0.06}$ &
                           $0.949$ & $0.9877 \pm 0.0041$
   \\
   $18\;-\;25$ &
             $0.713$ & $\pm 0.062$ & $^{+0.009}_{-0.085}$ & $^{+0.030}_{-0.029}$ &
                           $0.958$ & $0.9888 \pm 0.0031$
   \\
   $25\;-\;100$ &
             $0.0271$ & $\pm 0.0037$ & $^{+0.0011}_{-0.0001}$ & $^{+0.0012}_{-0.0013}$ &
                           $0.951$ & $0.9808 \pm 0.0035$
   \\
\hline
\hline
\multicolumn{7}{|c|}{\raisebox{0.05cm}[0.5cm]{$\mathbf{1000 < Q^2 < 2000 \text{ \bf GeV}^2}$}} \\
\hline
   $8\;-\;10$ &
              $2.80$  & $\pm 0.22$ & $^{+0.02}_{-0.06}$ & $^{+0.05}_{-0.07}$ &
                           $0.934$ & $0.9170 \pm 0.0078$
   \\
   $10\;-\;14$ & 
             $1.86$  & $\pm 0.14$ & $^{+0.05}_{-0.02}$ & $^{+0.03}_{-0.05}$ &
                           $0.937$ & $0.9567 \pm 0.0041$
   \\
   $14\;-\;18$ &
             $1.006$ & $\pm 0.099$ & $^{+0.080}_{-0.025}$ & $^{+0.029}_{-0.027}$ &
                           $0.945$ & $0.9856 \pm 0.0049$
   \\
   $18\;-\;25$ &
             $0.287$ & $\pm 0.037$ & $^{+0.022}_{-0.028}$ & $^{+0.011}_{-0.008}$ &
                           $0.936$ & $0.9976 \pm 0.0006$
   \\
   $25\;-\;100$ &
             $0.0173$ & $\pm 0.0030$ & $^{+0.0043}_{-0.0004}$ & $^{+0.0006}_{-0.0006}$ &
                           $0.943$ & $0.9946 \pm 0.0038$
   \\
\hline
\end{tabular}
\caption{
Inclusive jet cross-section $d\sigma/d\etjetb$ in different regions of $Q^2$ for 
jets of hadrons in the Breit frame,
selected with the longitudinally invariant $k_T$ cluster algorithm. For details, see the caption 
of Table~\ref{tablaq2}.
}
  \label{tablaetq21}
\end{center}
\end{table}
\begin{table}[p]
\begin{center}
\begin{tabular}{|c|cccc||c|c|}
\hline
\raisebox{0.25cm}[1.cm]{\parbox{2cm}{\centerline{$\etjetb$ bin} {\centerline{(GeV)}}}} & 
\raisebox{0.25cm}[1.cm]{\parbox{2cm}{\centerline{$d\sigma/d\etjetb$} \centerline{(pb/GeV)}}} & $\Delta_{stat}$ & 
                                     $\Delta_{syst}$ & 
                                     $\Delta_{\text{jet}-ES}$ & 
            \raisebox{0.2cm}[0.8cm]{\parbox{2.cm}{\centerline{QED} \vspace{-.2cm} \centerline{correction}}} &  
           \raisebox{0.2cm}[0.8cm]{\parbox{2.8cm}{\centerline{PAR to HAD} \vspace{-0.2cm} \centerline{correction}}}\\[.05cm]
\hline
\hline
\multicolumn{7}{|c|}{\raisebox{0.05cm}[0.5cm]{$\mathbf{2000 < Q^2 < 5000 \text{ \bf GeV}^2}$}} \\
\hline
   $8\;-\;10$ &
              $1.30$  & $\pm 0.15$ & $^{+0.08}_{-0.09}$ & $^{+0.01}_{-0.02}$ &
                           $0.934$ & $0.9143 \pm 0.0097$
   \\
   $10\;-\;14$ & 
             $0.724$  & $\pm 0.081$ & $^{+0.124}_{-0.048}$ & $^{+0.014}_{-0.015}$ &
                           $0.938$ & $0.9521 \pm 0.0058$
   \\
   $14\;-\;18$ &
             $0.318$ & $\pm 0.051$ & $^{+0.010}_{-0.029}$ & $^{+0.009}_{-0.008}$ &
                           $0.941$ & $0.9869 \pm 0.0076$
   \\
   $18\;-\;25$ &
             $0.209$ & $\pm 0.035$ & $^{+0.006}_{-0.041}$ & $^{+0.002}_{-0.002}$ &
                           $0.941$ & $0.9955 \pm 0.0016$
   \\
   $25\;-\;100$ &
             $0.0167$ & $\pm 0.0032$ & $^{+0.0018}_{-0.0044}$ & $^{+0.0007}_{-0.0007}$ &
                           $0.950$ & $1.0022 \pm 0.0039$
   \\
\hline
\hline
\multicolumn{7}{|c|}{\raisebox{0.05cm}[0.5cm]{$\mathbf{Q^2 > 5000 \text{ \bf GeV}^2}$}} \\
\hline
   $8\;-\;10$ &
              $0.258$  & $\pm 0.073$ & $^{+0.029}_{-0.028}$ & $^{+0.003}_{-0.009}$ &
                           $0.998$ & $0.940 \pm 0.017$
   \\
   $10\;-\;14$ & 
             $0.162$  & $\pm 0.042$ & $^{+0.007}_{-0.034}$ & $^{+0.007}_{-0.003}$ &
                           $0.934$ & $0.958 \pm 0.012$
   \\
   $14\;-\;18$ &
             $0.110$ & $\pm 0.032$ & $^{+0.003}_{-0.006}$ & $^{+0.000}_{-0.005}$ &
                           $0.937$ & $0.9777 \pm 0.0026$
   \\
   $18\;-\;25$ &
             $0.055$ & $\pm 0.018$ & $^{+0.015}_{-0.000}$ & $^{+0.001}_{-0.001}$ &
                           $0.936$ & $0.9994 \pm 0.0069$
   \\
   $25\;-\;100$ &
             $0.0036$ & $\pm 0.0014$ & $^{+0.0007}_{-0.0000}$ & $^{+0.0001}_{-0.0001}$ &
                           $0.927$ & $1.00291 \pm 0.00090$
   \\
\hline
\end{tabular}
\caption{
Continuation of Table~\ref{tablaetq21}.  For details, see the caption 
of Table~\ref{tablaq2}.
}
  \label{tablaetq22}
\end{center}
\end{table}
\begin{table}[p]
\begin{center}
\begin{tabular}{|c|cccc|}
\hline
\raisebox{0.25cm}[1.0cm]{\parbox{2cm}{\centerline{$Q^2$ region} \centerline{(GeV$^2$)}}}
            & \raisebox{0.15cm}[1.cm]{\parbox{2cm}{\centerline{$\alpha_s (M_Z)$}}} 
                           & $\Delta_{stat}$ & 
                             $\Delta_{syst}$ & 
                             $\Delta_{th}$ \\[0.05cm]
\hline
   $125\;-\;250$ &
              $0.1252$  & $\pm 0.0013$ &  $^{+0.0042}_{-0.0048}$ &  $^{+0.0082}_{-0.0062}$ 
   \\[.1cm]
   $250\;-\;500$ & 
              $0.1264$  &  $\pm 0.0017$ & $^{+0.0036}_{-0.0047}$ & $^{+0.0060}_{-0.0036}$ 
   \\[.1cm]
   $500\;-\;1000$ &
              $0.1203$ &  $^{+0.0022}_{-0.0023}$ & $^{+0.0027}_{-0.0038}$ & $^{+0.0032}_{-0.0016}$ 
   \\[.1cm]
   $1000\;-\;2000$ &
              $0.1208$ &  $\pm 0.0032$ & $^{+0.0025}_{-0.0021}$ & $^{+0.0022}_{-0.0022}$ 
   \\[.1cm]
   $2000\;-\;5000$ &
              $0.1256$ &  $^{+0.0047}_{-0.0049}$ & $^{+0.0045}_{-0.0079}$ & $^{+0.0040}_{-0.0041}$
   \\[.1cm]
   $5000\;-\;10^{5}$ &
              $0.1286$ & $^{+0.0146}_{-0.0158}$ & $^{+0.0055}_{-0.0046}$ & $^{+0.0045}_{-0.0044}$ 
   \\
\hline
   $ > 125$ &
              $0.1244$ &  $\pm 0.0009$ & $^{+0.0034}_{-0.0041}$ & $^{+0.0057}_{-0.0040}$ 
   \\[.1cm]
   $ > 500$ &
              $0.1212$ & $\pm 0.0017$ & $^{+0.0023}_{-0.0031}$ & $^{+0.0028}_{-0.0027}$ 
   \\
\hline
\end{tabular}
\caption{
The $\alpha_s (M_Z)$ values 
as determined from the QCD fit to the
measured $d\sigma/dQ^2$, as well as 
those obtained by combining several regions in that distribution. The statistical, systematic and 
theoretical uncertainties are shown separately.
}
  \label{tablaasq2}
\end{center}
\end{table}
\begin{table}[p]
\begin{center}
\begin{tabular}{|c|cccc|}
\hline
\raisebox{0.25cm}[1.cm]{\parbox{2cm}{\centerline{$\etjetb$ region} \centerline{(GeV)}}}
            & \raisebox{0.15cm}[1.cm]{\parbox{2cm}{\centerline{$\alpha_s (M_Z)$}}} 
                           & $\Delta_{stat}$ & 
                             $\Delta_{syst}$ & 
                             $\Delta_{th}$ \\[0.05cm]
\hline
   $8\;-\;10$ &
              $0.1285$  & $\pm 0.0015$ &  $^{+0.0038}_{-0.0052}$ &  $^{+0.0078}_{-0.0046}$ 
   \\[.1cm]
   $10\;-\;14$ & 
              $0.1238$  &  $\pm 0.0013$ & $^{+0.0037}_{-0.0036}$ & $^{+0.0060}_{-0.0042}$ 
   \\[.1cm]
   $14\;-\;18$ &
              $0.1236$ &  $\pm 0.0017$ & $^{+0.0030}_{-0.0039}$ & $^{+0.0046}_{-0.0035}$ 
   \\[.1cm]
   $18\;-\;25$ &
              $0.1188$ &  $^{+0.0021}_{-0.0022}$ & $^{+0.0032}_{-0.0037}$ & $^{+0.0036}_{-0.0026}$ 
   \\[.1cm]
   $25\;-\;35$ &
              $0.1157$ &  $^{+0.0043}_{-0.0044}$ & $^{+0.0029}_{-0.0027}$ & $^{+0.0039}_{-0.0029}$ 
   \\[.1cm]
   $35\;-\;100$ &
              $0.1422$ & $^{+0.0174}_{-0.0178}$ & $^{+0.0096}_{-0.0112}$ & $^{+0.0106}_{-0.0088}$ 
   \\
\hline
   $ > 8 $ &
              $0.1241$  & $\pm 0.0008$ &  $^{+0.0034}_{-0.0039}$ &  $^{+0.0055}_{-0.0038}$ 
   \\[.1cm]
   $ >14 $ & 
              $0.1212$  &  $\pm 0.0013$ & $^{+0.0030}_{-0.0036}$ & $^{+0.0041}_{-0.0030}$ 
   \\
\hline
\end{tabular}
\caption{
The $\alpha_s (M_Z)$ values 
as determined from the QCD fit to the
measured $d\sigma/d\etjetb$, as well as  
those obtained by combining several regions in that distribution. The statistical, systematic and 
theoretical uncertainties are shown separately.
}
  \label{tablaaset}
\end{center}
\end{table}
\begin{table}[p]
\begin{center}
\begin{tabular}{|c|cccc|}
\hline
\raisebox{0.25cm}[1.cm]{\parbox{2cm}{\centerline{\big<$\etjetb$\big>} \centerline{(GeV)}}}
            & \raisebox{0.15cm}[1.cm]{\parbox{2.5cm}{\centerline{$\alpha_s (\big<\etjetb\big>)$}}} 
                           & $\Delta_{stat}$ & 
                             $\Delta_{syst}$ & 
                             $\Delta_{th}$ \\[0.05cm]
\hline
   $8.91$ &
              $0.2113$  & $^{+0.0042}_{-0.0041}$ &  $^{+0.0109}_{-0.0144}$ &  $^{+0.0228}_{-0.0127}$ 
   \\[.1cm]
   $11.65$ & 
              $0.1851$  &  $\pm 0.0030$ & $^{+0.0087}_{-0.0083}$ & $^{+0.0141}_{-0.0095}$ 
   \\[.1cm]
   $15.70$ &
              $0.1721$ &  $\pm 0.0034$ & $^{+0.0059}_{-0.0078}$ & $^{+0.0094}_{-0.0068}$ 
   \\[.1cm]
   $20.69$ &
              $0.1538$ &  $^{+0.0036}_{-0.0037}$ & $^{+0.0054}_{-0.0063}$ & $^{+0.0062}_{-0.0044}$ 
   \\[.1cm]
   $28.61$ &
              $0.1398$ &  $^{+0.0064}_{-0.0065}$ & $^{+0.0044}_{-0.0040}$ & $^{+0.0057}_{-0.0043}$ 
   \\[.1cm]
   $41.98$ &
              $0.1660$ & $^{+0.0247}_{-0.0240}$ & $^{+0.0135}_{-0.0153}$ & $^{+0.0148}_{-0.0121}$ 
   \\
\hline
\end{tabular}
\caption{
The $\alpha_s(\big<\etjetb\big>)$ values 
as determined from the QCD fit to the
measured $d\sigma/d\etjetb$.
The statistical, systematic and 
theoretical uncertainties are shown separately.
}
  \label{tablaaset2}
\end{center}
\end{table}
%
%

%
%
\end{document}